\documentclass[english,prl, two column, superscriptaddress]{revtex4-1}
\usepackage[T1]{fontenc}
\usepackage[latin9]{inputenc}
\usepackage{color}
\usepackage{babel}
\usepackage{amssymb}
\usepackage{stackrel}
\usepackage{graphicx}
\usepackage[unicode=true,pdfusetitle,
 bookmarks=true,bookmarksnumbered=false,bookmarksopen=false,
 breaklinks=false,pdfborder={0 0 0},backref=false,colorlinks=true]
 {hyperref}
\hypersetup{
 colorlinks,linkcolor=red,citecolor=blue}
\usepackage{breakurl}

\makeatletter

\newcommand{\lyxdot}{.}

\@ifundefined{textcolor}{}
{%
 \definecolor{BLACK}{gray}{0}
 \definecolor{WHITE}{gray}{1}
 \definecolor{RED}{rgb}{1,0,0}
 \definecolor{GREEN}{rgb}{0,1,0}
 \definecolor{BLUE}{rgb}{0,0,1}
 \definecolor{CYAN}{cmyk}{1,0,0,0}
 \definecolor{MAGENTA}{cmyk}{0,1,0,0}
 \definecolor{YELLOW}{cmyk}{0,0,1,0}
}

\usepackage{braket}

\makeatother

\begin{document}

\title{Cat codes with optimal decoherence suppression for a lossy bosonic
channel}

\author{Linshu Li }

\affiliation{Departments of Applied Physics and Physics, Yale University, New
Haven, CT 06511, USA}

\author{Chang-ling Zou}

\affiliation{Departments of Applied Physics and Physics, Yale University, New
Haven, CT 06511, USA}

\author{Victor V. Albert}

\affiliation{Departments of Applied Physics and Physics, Yale University, New
Haven, CT 06511, USA}

\author{Sreraman Muralidharan}

\affiliation{Department of Electrical Engineering, Yale University, New Haven,
CT 06511, USA}

\author{S. M. Girvin}

\affiliation{Departments of Applied Physics and Physics, Yale University, New
Haven, CT 06511, USA}

\author{Liang Jiang}

\affiliation{Departments of Applied Physics and Physics, Yale University, New
Haven, CT 06511, USA}
\begin{abstract}
We investigate cat codes that can correct multiple excitation losses
and identify two types of logical errors: bit-flip errors due to excessive
excitation loss and dephasing errors due to quantum back-action from
the environment. We show that selected choices of logical subspace
and coherent amplitude can efficiently reduce dephasing errors. The
trade-off between the two major errors enables optimized performance
of cat codes in terms of minimized decoherence. With high coupling
efficiency, we show that one-way quantum repeaters with cat codes
feature drastically boosted secure communication rate per mode compared
with conventional encoding schemes, and thus showcase the promising
potential of quantum information processing with continuous variable
quantum codes.
\end{abstract}
\maketitle
An outstanding challenge for quantum information processing with bosonic
systems is excitation loss, which can be modeled as a lossy bosonic
channel (LBC) \cite{Chuang1997,Cochrane1999}. To suppress excitation
loss, the conventional approach is to consider discrete variable (DV)
encodings that use physical qubits (qudits) implemented with a single
excitation distributed over two (multiple) bosonic modes and standard
qubit- (qudit-) based quantum error correction (QEC) \cite{Ralph2005,Varnava2006,Varnava2007}.
Such DV encoding schemes usually require a considerable number of
bosonic modes to encode one logical qubit (qudit). In contrast, continuous
variable (CV) encoding schemes deploy the Hilbert space of higher
excitations, enabling single-mode based QEC against loss errors. The
resulting mode-efficiency can potentially lead to high storage-density
quantum memories and boost the secure communication rate per mode
for long distance quantum communication \cite{Lo2014,Takeoka2014,Pirandola2015,Briegel1998,Jiang2009,Munro2012,Muralidharan2015}. 

Cat codes \cite{Cochrane1999,Leghtas2013,Mirrahimi2014}, among other
single-mode CV schemes \cite{Gottesman2001,Michael2016}, have been
proposed for correcting excitation loss. With the rapid development
of quantum control \cite{Krastanov2015,Heeres2015} and high-fidelity
quantum non-demolition readout \cite{Murch2013,Hatridge2013,Sun2014},
QEC with cat codes has recently been demonstrated to reach the break-even
point in superconducting circuits \cite{Ofek2016}. These advances
have opened up a new era of CV quantum information in which states
can be stored and manipulated for a duration longer than the intrinsic
coherence time of the constituent modes.

Cat codes are based on coherent superpositions of coherent states.
Qualitatively it has been known that a proper choice of coherent amplitude
$\alpha$ is essential for QEC with cat codes: A large $\alpha$ increases
the probability of uncorrectable excitation loss while a small $\alpha$
may lead to significant overlap between neighboring coherent components.
Yet, to date, the optimal choice of $\alpha$ and hence the optimal
QEC capability of cat codes has remained unquantified. In this letter,
we investigate cat codes that encode a logical qubit using superpositions
of $2d$ coherent components and can correct up to $d-1$ excitation
losses \cite{Leghtas2013,Mirrahimi2014}. We quantify the two major
types of errors associated with the encoding: the logical bit-flip
errors due to imperfect capability of correcting excitation loss,
and the logical dephasing errors induced by back-action from the environment.
The analysis allows us to find non-trivial choices of code parameters
that significantly reduce the back-action and also balance the two
logical errors. Using parameters that yield minimum decoherence, we
analyze the performance of cat codes in one-way quantum repeaters
(QRs) for ultrafast quantum communication over transcontinental scales.

\paragraph*{Lossy bosonic channel.}

The Kraus operator-sum representation for the LBC is \cite{Chuang1997}
\begin{equation}
\mathcal{L}\left(\rho\right)=\stackrel[k=0]{\infty}{\sum}\hat{\mathcal{W}}_{k}\rho\hat{\mathcal{W}}_{k}^{\dagger},
\end{equation}
where $\hat{\mathcal{W}}_{k}=\frac{1}{\sqrt{k!}}\gamma^{\frac{k}{2}}\left(1-\gamma\right)^{a^{\dagger}a/2}a^{k}$
is the Kraus operator associated with losing $k$ excitations, $a$
($a^{\dagger}$) is the boson annihilation (creation) operator, and
$\gamma$ is the loss probability of each excitation. Excitation loss
in bosonic systems, such as localized cavity modes for quantum memories
and propagating modes for quantum communication, can be modeled as
a LBC. For cavities, $\gamma=1-e^{-\kappa t}$, where $\kappa$ is
the decay constant and $t$ is the storage time; for propagating modes
with attenuation length $L_{\mathrm{att}}$, $\gamma=1-\eta^{2}e^{-L/L_{\mathrm{att}}}$,
where $L$ is the propagation distance and $\eta$ is the coupling
efficiency of the interface between the optical fiber and local processing
devices.

\paragraph*{Cat codes and properties.}

\begin{figure}[h]
\begin{raggedright}
\includegraphics[scale=0.43]{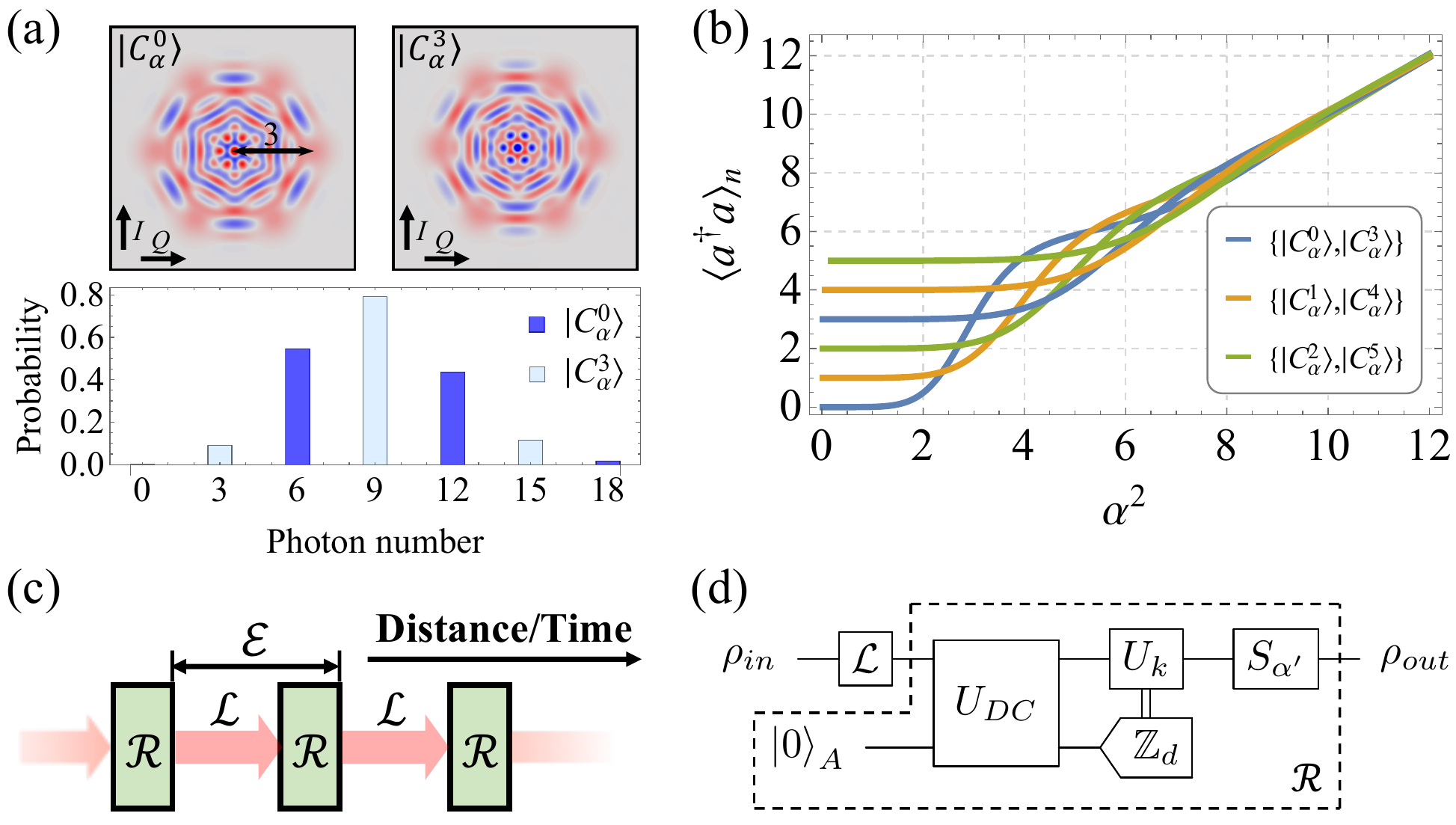}
\par\end{raggedright}

\protect\caption{(a) Wigner functions and excitation number distributions of $\left|C_{\alpha}^{0}\right\rangle $
and $\left|C_{\alpha}^{3}\right\rangle $ with $d=\alpha=3$. (b)
Average excitation number $\left\langle a^{\dagger}a\right\rangle _{n}$
for cat states with $d=3$. (c) Schematic of alternating LBC ($\mathcal{L}$)
and QEC recovery ($\mathcal{R}$). (d) Quantum circuits of QEC recovery
for cat codes, consisting of the dispersive coupling gate $U_{DC}$
followed by $\mathbb{Z}_{d}$ measurement of excitation number, rotation
gate $U_{k}$ conditioned on $\mathbb{Z}_{d}$ measurement outcome
to compensates the lost excitations, and finally amplitude restoration
$S_{\alpha'}$. \label{fig:Cat states and circuit} }
\end{figure}

The basis states of cat codes are superpositions of coherent states
lying equidistantly on a circle in the phase space of a single bosonic
mode. We define the orthonormal basis associated with $2d$ coherent
states 
\begin{eqnarray}
\left|C_{\alpha}^{n}\right\rangle  & = & \frac{1}{\sqrt{2d\mathcal{N}_{n}\left(\alpha\right)}}\stackrel[k=0]{2d-1}{\sum}\omega^{-kn}\left|\omega^{k}\alpha\right\rangle ,
\end{eqnarray}
 where $\omega=e^{i\frac{\pi}{d}}$ and $\mathcal{N}_{n}\left(\alpha\right)=\stackrel[k=0]{2d-1}{\sum}\omega^{-kn}e^{\left(\omega^{k}-1\right)\alpha^{2}}$
is the normalization factor for $n=0,1,2,\cdots,2d-1$ \cite{Albert2014,Albert2016}.
Without losing generality, we assume $\alpha$ is real and positive.
Since each cat state $\left|C_{\alpha}^{n}\right\rangle $ is a superposition
of $n\bmod2d$ number states ($\left|n\right\rangle $, $\left|n+2d\right\rangle $,
$\left|n+4d\right\rangle $,...), cat states are orthonormal ($\Braket{C_{\alpha}^{n_{1}}|C_{\alpha}^{n_{2}}}=\delta_{n_{1}n_{2}}$).
The average excitation number $\left\langle C_{\alpha}^{n}\right|a^{\dagger}a\left|C_{\alpha}^{n}\right\rangle =\alpha^{2}\mathcal{N}_{n-1}\left(\alpha\right)/\mathcal{N}_{n}\left(\alpha\right)\rightarrow\alpha^{2}$
for $\alpha\rightarrow\infty$ \cite{SupplementalMaterial}, as shown
in Fig.~\ref{fig:Cat states and circuit}(b), while for finite $\alpha$
it deviates from $\alpha^{2}$ due to the oscillatory $\mathcal{N}_{n-1}\left(\alpha\right)/\mathcal{N}_{n}\left(\alpha\right)$.

The $2d$-dimensional cat Hilbert space can be divided into $d$ subspaces
labeled by $s=0,1,\cdots,d-1$. The ``$s$-subspace'' has excitation
number $s\bmod\ d$, spanned by two logical states $\left|0\right\rangle _{L}^{s}=\left|C_{\alpha}^{s}\right\rangle $
and $\left|1\right\rangle _{L}^{s}=\left|C_{\alpha}^{s+d}\right\rangle $.
Fig.~\ref{fig:Cat states and circuit}(a) shows the Wigner functions
and excitation distributions of $\left|C_{\alpha}^{s}\right\rangle $
and $\left|C_{\alpha}^{s+d}\right\rangle $ for $d=3$, $\alpha=3$
and $s=0$. It becomes clear that $d$, $\alpha$, and $s$ are three
degrees of freedom that determine the performance of cat codes in
correcting loss errors.

After losing $k$ excitations, the $s$-subspace is mapped to the
$\left(s-k\right)$-subspace: $\left|C_{\alpha}^{s}\right\rangle \rightarrow\left|C_{\alpha}^{s-k}\right\rangle $
and $\left|C_{\alpha}^{s+d}\right\rangle \rightarrow\left|C_{\alpha}^{s+d-k}\right\rangle $
. Hence, we can unambiguously distinguish $0\le k\le d-1$ excitation
losses without destroying the encoded logical states by projectively
measuring the excitation number $\mathrm{mod}$ $d$ (called ``$\mathbb{Z}_{d}$
measurement''). In fact, since a cat state maps back to itself after
losing integer multiples of $2d$ excitations, we can restore the
logical basis states correctly with $2md\le k\le\left(2m+1\right)d-1$
excitation losses for integer $m$. If there are $\left(2m+1\right)d\le k\le2\left(m+1\right)d-1$
excitation losses, however, we will misidentify the logical basis
states. Since the symmetric superposition $\left|C_{\alpha}^{s}\right\rangle +\left|C_{\alpha}^{s+d}\right\rangle \rightarrow\left|C_{\alpha}^{s-k}\right\rangle +\left|C_{\alpha}^{s+d-k}\right\rangle $
is actually preserved even if we misidentify the logical basis, the
misidentification effectively induces an X rotation in the logical
basis -- a\textit{ logical bit-flip error}.

In addition to the logical bit-flip error, the LBC can induce another
type of error via back-action from the environment. For finite $\alpha$,
the logical states $\left|C_{\alpha}^{s}\right\rangle $ and $\left|C_{\alpha}^{s+d}\right\rangle $
generally differ in average photon number, as illustrated in Fig.~\ref{fig:Cat states and circuit}(b),
as well as the $m$-th moments $\left\langle C_{\alpha}^{s}\right|\left(a^{\dagger}a\right)^{m}\left|C_{\alpha}^{s}\right\rangle \neq\left\langle C_{\alpha}^{d+s}\right|\left(a^{\dagger}a\right)^{m}\left|C_{\alpha}^{d+s}\right\rangle $
for $m\in\mathbb{Z}^{+}$. Hence, the excitation loss to the environment
can leak out information about the logical state, which is captured
by Kraus operator acting on logical states, $\hat{\mathcal{W}}_{k}\left|C_{\alpha}^{n}\right\rangle \propto\left(1-\gamma\right)^{a^{\dagger}a/2}a^{k}\left|C_{\alpha}^{n}\right\rangle =e^{-\Delta}\alpha^{k}\sqrt{\mathcal{N}_{n-k}\left(\alpha'\right)/\mathcal{N}_{n}\left(\alpha\right)}\left|C_{\alpha'}^{n-k}\right\rangle $
with $\alpha'=\sqrt{1-\gamma}\alpha$ and $\Delta=\alpha^{2}-\alpha'^{2}=\gamma\alpha^{2}$.
Defining $G\left(n,m\right)=\sqrt{\mathcal{N}_{m}\left(\alpha'\right)/\mathcal{N}_{n}\left(\alpha\right)}$,
the fact that $G\left(n,n-k\right)$ is slightly different for $n=s$
and $n=s+d$ results in the back-action associated with losing $k$
excitations %
\footnote{The back-action also corresponds to the subtle bias from likelihood
estimate of the logical states after the $\mathbb{Z}_{d}$ measurement.%
}. When we average over all possible $k$ values, the back-action induced
bias towards $\left|C_{\alpha}^{s}\right\rangle $ or $\left|C_{\alpha}^{s+d}\right\rangle $
are mostly cancelled. However, the back-action does reduce the coherence
between $\left|C_{\alpha}^{s}\right\rangle $ and $\left|C_{\alpha}^{s+d}\right\rangle $
and effectively induces a \textit{logical dephasing error}.

\paragraph*{QEC recovery for cat codes.}

To protect the quantum information from bosonic loss, we introduce
a QEC recovery operation $\mathcal{R}$ (shown in Fig.~\ref{fig:Cat states and circuit}(d)),
which consists of a $\mathbb{Z}_{d}$ measurement, conditional loss
compensation, and amplitude restoration. First, we use the $\mathbb{Z}_{d}$
measurement to distinguish different loss events up to losing $d-1$
excitations. Similar to the qubit-assisted number parity ($\mathbb{Z}_{2}$)
measurement \cite{Sun2014}, we consider a $d$-level ancilla (e.g.,
using higher levels of the transmon \cite{Wang2016}) that dispersively
couples to a cavity mode 
\begin{equation}
\hat{H}_{DC}=\stackrel[j=0]{d-1}{\sum}j\chi\left|j\right\rangle \left\langle j\right|a^{\dagger}a,\label{eq:dispersive coupling hamiltonian}
\end{equation}
where $\left|j\right\rangle $ are the basis states of the ancilla.
Combined with Fourier gates on the $d$-level ancilla, $F_{d}$, we
can implement the unitary operation $U_{DC}=F_{d}^{\dagger}e^{-i\frac{\pi}{\chi}H_{DC}}F_{d}$
that maps the $\mathbb{Z}_{d}$ information to the ancilla that is
subsequently measured in $\left\{ \left|j\right\rangle \right\} $
basis. 

Then, conditioned on measured excitation loss number (mod $d$), $k\in\left\{ 0,1,\cdots,d-1\right\} $,
we implement the following unitary to compensate the identified loss
and restore the state back to the $s$-subspace 
\begin{equation}
U_{k}=\left|C_{\alpha'}^{s}\right\rangle \left\langle C_{\alpha'}^{s-k}\right|+\left|C_{\alpha'}^{d+s}\right\rangle \left\langle C_{\alpha'}^{d+s-k}\right|+U_{k}^{0},\label{eq: photon-addition mapping}
\end{equation}
where $U_{k}^{0}$ is an arbitrary unitary on the complementary subspace
of $\left\{ \left|C_{\alpha'}^{s-k}\right\rangle ,\left|C_{\alpha'}^{d+s-k}\right\rangle \right\} $
so that $U_{k}$ is a unitary in the entire Hilbert space. $U_{k}$
can be achieved with unitary control of the bosonic mode (e.g., as
demonstrated in superconducting circuits \cite{Krastanov2015,Heeres2015,Heeres2016}). 

Finally, we restore the amplitude from $\alpha'$ back to $\alpha$
via the following unitary 
\begin{equation}
S_{\alpha'}=\sum_{s=0}^{d-1}\left(\left|C_{\alpha}^{s}\right\rangle \left\langle C_{\alpha'}^{s}\right|+\left|C_{\alpha}^{d+s}\right\rangle \left\langle C_{\alpha'}^{d+s}\right|\right)+S_{\alpha'}^{0},\label{eq: Amplitude repumping}
\end{equation}
where $S_{\alpha'}^{0}$ is an arbitrary unitary on the complementary
subspace of the $2d$-dimensional subspace spanned by $\left|C_{\alpha'}^{n}\right\rangle $.
Alternative to the unitary implementation of $S_{\alpha'}$, we may
also use engineered dissipation to restore the amplitude from $\alpha'$
to $\alpha$ without compromising the encoded logical state \cite{Leghtas2015,Mirrahimi2014}.

Overall, the QEC recovery in Fig.~\ref{fig:Cat states and circuit}(d)
implements
\begin{equation}
\mathcal{R}\left(\rho\right)=\stackrel[k=0]{d-1}{\sum}\left|C_{\alpha}^{s}\right\rangle \left\langle C_{\alpha'}^{s-k}\right|\rho\left|C_{\alpha'}^{s-k}\right\rangle \left\langle C_{\alpha}^{s}\right|,
\end{equation}
which restores the original encoded subspace. Note that the $d$-level
ancilla can also be replaced by a $2$-level ancilla, with an overhead
of $\log_{2}d$ steps of measurement and feedforward control to fully
implement the QEC recovery $\mathcal{R}$ with Kraus rank $d$ \cite{Lloyd2001,Shen}.

\paragraph*{Logical bit-flip and dephasing errors.}

We now analyze the effective errors in the encoded subspace after
the QEC recovery. Writing density matrix as a column vector $\rho=\left(\begin{array}{cccc}
\rho_{00} & \rho_{01} & \rho_{10} & \rho_{11}\end{array}\right)^{\mathrm{T}}$, $\rho_{out}$ is linked to $\rho_{in}$ via $\mathcal{E}=\mathcal{R}\circ\mathcal{L}=\left(\mathcal{E}_{ij}\right)\left(i,j=1,2,3,4\right)$
\begin{eqnarray}
\mathcal{E} & = & \stackrel[k=0]{d-1}{\sum}T_{k}\left(\begin{array}{cccc}
A_{sk}^{2} & 0 & 0 & 0\\
0 & A_{sk}D_{sk} & 0 & 0\\
0 & 0 & A_{sk}D_{sk} & 0\\
0 & 0 & 0 & D_{sk}^{2}
\end{array}\right)\nonumber \\
 &  & +\stackrel[k=0]{d-1}{\sum}T_{k+d}\left(\begin{array}{cccc}
0 & 0 & 0 & C_{sk}^{2}\\
0 & 0 & B_{sk}C_{sk} & 0\\
0 & B_{sk}C_{sk} & 0 & 0\\
B_{sk}^{2} & 0 & 0 & 0
\end{array}\right),\label{eq: quantum channel}
\end{eqnarray}
where $A_{sk}=G\left(s,s-k\right)$, $B_{sk}=G\left(s,d+s-k\right)$,
$C_{sk}=G\left(d+s,s-k\right)$ and $D_{sk}=G\left(d+s,d+s-k\right)$
are back-action coefficients. $T_{k}=\stackrel[m=0]{\infty}{\sum}\frac{e^{-\Delta}\Delta^{2md+k}}{\left(2md+k\right)!}$
and $T_{k+d}=\stackrel[m=0]{\infty}{\sum}\frac{e^{-\Delta}\Delta^{\left(2m+1\right)d+k}}{\left[\left(2m+1\right)d+k\right]!}$
is the probability for correct and incorrect recovery, respectively,
for an ideal Poisson distribution with mean $\Delta$ for excitation
losses. It is clear from Eq.~(\ref{eq: quantum channel}) that the
probabilities for excitation losses for cat codes are modulated by
the back-action.
\begin{figure}[h]
\begin{centering}
\includegraphics[scale=0.7]{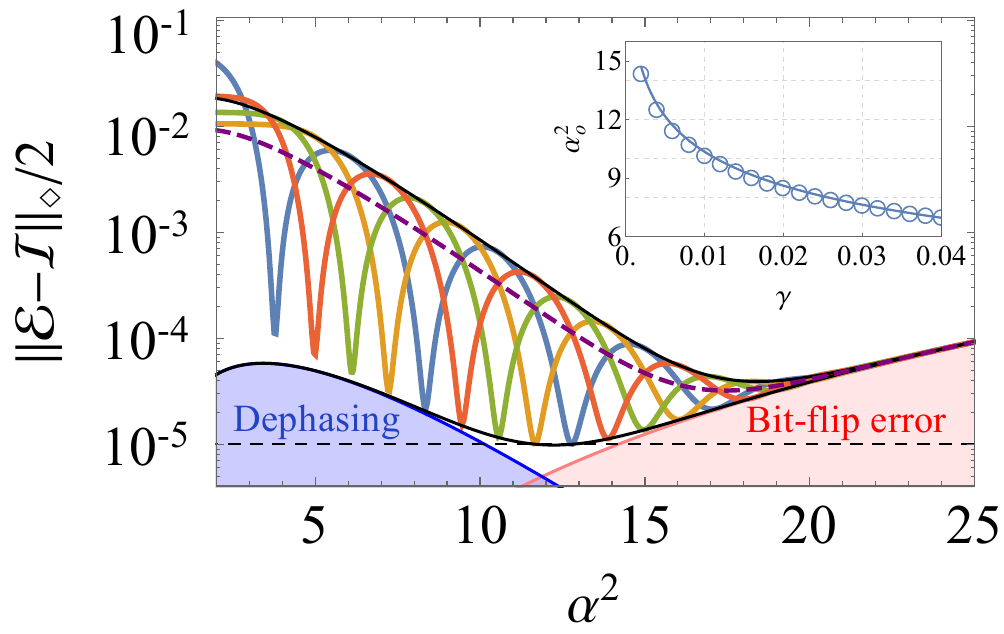} 
\par\end{centering}

\raggedright{}\protect\caption{Diamond distance $\left\Vert \mathcal{E}-\mathcal{I}\right\Vert _{\diamond}/2$
calculated numerically from the $\mathcal{E}$ in Eq.~(\ref{eq: quantum channel})
for logical subspace $s=0,1,2,3$ (blue, red, green and yellow curves,
respectively) and analytical bounds $\Gamma_{\pm}$ (black), for $d=4$
and $\gamma=0.005$. The two types of errors in $\Gamma_{-}$, i.e.
logical bit-flip error $\epsilon_{f}$ and logical dephasing $\epsilon_{d}$,
are marked. The dashed purple and black curves show $\bar{\Gamma}$
and the minimum $\Gamma_{-}$, respectively. The inset shows $\alpha_{o}^{2}\left(\gamma\right)$
for $d=4$; analytical results (solid) from Eq.~(\ref{eq:optimum asq})
agree with numerical calculations (circles).\label{analytical v.s. numerical}}
\end{figure}

Considering small logical bit-flip error and overlap between neighboring
coherent states, to the leading order of error we can write $\mathcal{E}$
as a Pauli channel \cite{SupplementalMaterial} 
\begin{equation}
\mathcal{E}\left(\rho\right)\approx\left[1-\left(\epsilon_{f}+\epsilon_{d}\right)\right]\rho+\epsilon_{f}X\rho X+\epsilon_{d}Z\rho Z,\label{eq:Pauli Channel}
\end{equation}
 with logical bit-flip error due to excessive loss of (more than $d-1$)
excitations 
\begin{eqnarray}
\epsilon_{f} & = & \stackrel[k=0]{d-1}{\sum}T_{k+d},\label{eq:bit-flip error}
\end{eqnarray}
and logical dephasing error induced by back-action
\begin{eqnarray}
\epsilon_{d} & = & \frac{1}{2}e^{-4\alpha^{2}\sin^{2}\frac{\pi}{2d}}\left(e^{4\Delta\sin^{2}\frac{\pi}{2d}}-1\right)\nonumber \\
 &  & -\frac{1}{2}e^{-4\alpha^{2}\sin^{2}\frac{\pi}{2d}}\sqrt{1-2e^{\mu}\cos\psi+e^{2\mu}}\cos\theta,\label{eq: back-action induced dephasing}
\end{eqnarray}
where $\mu=2\Delta\left(2\sin^{2}\frac{\pi}{2d}-\sin^{2}\frac{\pi}{d}\right)$,
$\psi=\Delta(2\sin\frac{\pi}{d}-\sin\frac{2\pi}{d})$, $\theta=\frac{2s\pi}{d}-2\alpha^{2}\sin\frac{\pi}{d}+\arctan\frac{e^{\mu}\sin\psi}{1-e^{\mu}\cos\psi}$.
With Eq.~(\ref{eq: quantum channel}), we can quantify the residual
decoherence after the QEC recovery using the diamond distance \cite{Sacchi2005,Benenti2010}
\begin{equation}
\Gamma\left(\alpha,\ d,\ \gamma,\ s\right)\equiv\left\Vert \mathcal{E}-\mathcal{I}\right\Vert _{\diamond}/2\approx\epsilon_{f}+\epsilon_{d}.
\end{equation}

For given $\gamma$ and $d$, we may select coherent amplitude $\alpha$
and logical subspace $s$ to minimize $\left\Vert \mathcal{E}-\mathcal{I}\right\Vert _{\diamond}/2$.
As illustrated in Fig.~\ref{analytical v.s. numerical}, for each
fixed $s$-subspace encoding, the diamond distance oscillates with
$\alpha^{2}$ and there is a set of $\alpha$ where the back-action
induced dephasing reaches a local minimum, suppressed to $\mathcal{O}\left[\left(\Delta\pi^{2}/d^{2}\right)^{2}\right]$
\cite{SupplementalMaterial}. In fact, each favorable combination
of $s$ and $\alpha$ gives the same average excitation number for
the logical states, $\left\langle C_{\alpha}^{s}\right|a^{\dagger}a\left|C_{\alpha}^{s}\right\rangle =\left\langle C_{\alpha}^{d+s}\right|a^{\dagger}a\left|C_{\alpha}^{d+s}\right\rangle $
(associated with the crossing points in Fig.~\ref{fig:Cat states and circuit}(b)),
while the residual back-action only comes from the difference in second
and higher moments of $a^{\dagger}a$. 

To estimate the minimum achievable error, we obtain analytical expressions
for two approximate envelop functions 
\begin{eqnarray}
\Gamma_{\pm}\left(\alpha,\ \gamma,\ d\right) & = & \epsilon_{f}+\left.\epsilon_{d}\right|_{\cos\theta=\mp1},\label{eq:Analytical bounds}
\end{eqnarray}
which provide upper and lower bounds on the diamond distance $\Gamma\left(\alpha,\ d,\ \gamma,\ s\right)$
for all $s$. As illustrated in Fig.~\ref{analytical v.s. numerical},
to achieve the minimum error $\Gamma_{-}$ (lower black curve), it
is crucial to perform combined optimization of $\alpha$ and $s$.
If we are non-selective in the logical subspace (i.e., averaging over
all $s$) and only optimize the coherent amplitude $\alpha$, the
averaged error is approximately $\Gamma_{+}/2$ (dashed purple curve),
which can be an order of magnitude larger than $\Gamma_{-}$ for the
parameter region of interest.  Moreover, the combined optimization
also leads to a smaller optimized coherent amplitude.

Using Eq.~(\ref{eq:Analytical bounds}), we can estimate the optimal
amplitude $\alpha_{o}$ by requiring the two competing errors be equal
in $\Gamma_{-}$. For $\Delta\ll d$, we obtain the following approximate
expression for $\alpha_{o}^{2}\left(d,\gamma\right)$ for $d>2$ 
\begin{eqnarray}
\alpha_{o}^{2} & \approx & \frac{W\left[\frac{4\sin^{2}\left(\frac{\pi}{2d}\right)-\gamma}{\left(d-2\right)\gamma}\left(\frac{d!}{2}\frac{\pi^{4}}{d^{4}}\right)^{\frac{1}{d-2}}\right]}{\frac{4\sin^{2}\left(\frac{\pi}{2d}\right)-\gamma}{d-2}},\label{eq:optimum asq}
\end{eqnarray}
where $W$ is the Lambert W function $z=f^{-1}\left(ze^{z}\right)=W\left(ze^{z}\right)$.
The inset of Fig.~\ref{analytical v.s. numerical} shows that even
at $d=4$, there is already good agreement between Eq.~(\ref{eq:optimum asq})
and numerical results. Based on the estimated $\alpha_{o}$, we can
identify the best combination of $\alpha^{*}$ and $s^{*}$ near the
vicinity of the minimum $\Gamma_{-}$.

\paragraph*{Application to repetitive correction.}

So far we have considered the performance of cat codes for a single
round of LBC followed by QEC recovery, and have identified the optimal
amplitude $\alpha_{o}$ and logical subspace $s$ for given $d$ and
$\gamma$. For practical applications, however, we may use multiple
rounds of LBC and QEC recovery, and optimize the frequency of recovery
to best maintain the coherence. In the following, we consider one-way
QRs with cat codes \cite{Munro2012,Muralidharan2014} over transcontinental
distances ($\geq10^{3}\mathrm{km}$). We note that the effect of localized
gates that induce photon loss can be treated similarly as coupling
inefficiency and thus the results obtained below for one-way QRs are
naturally applicable to localized repetitive QEC with leaky gates.

\begin{figure}[h]
\includegraphics[scale=0.44]{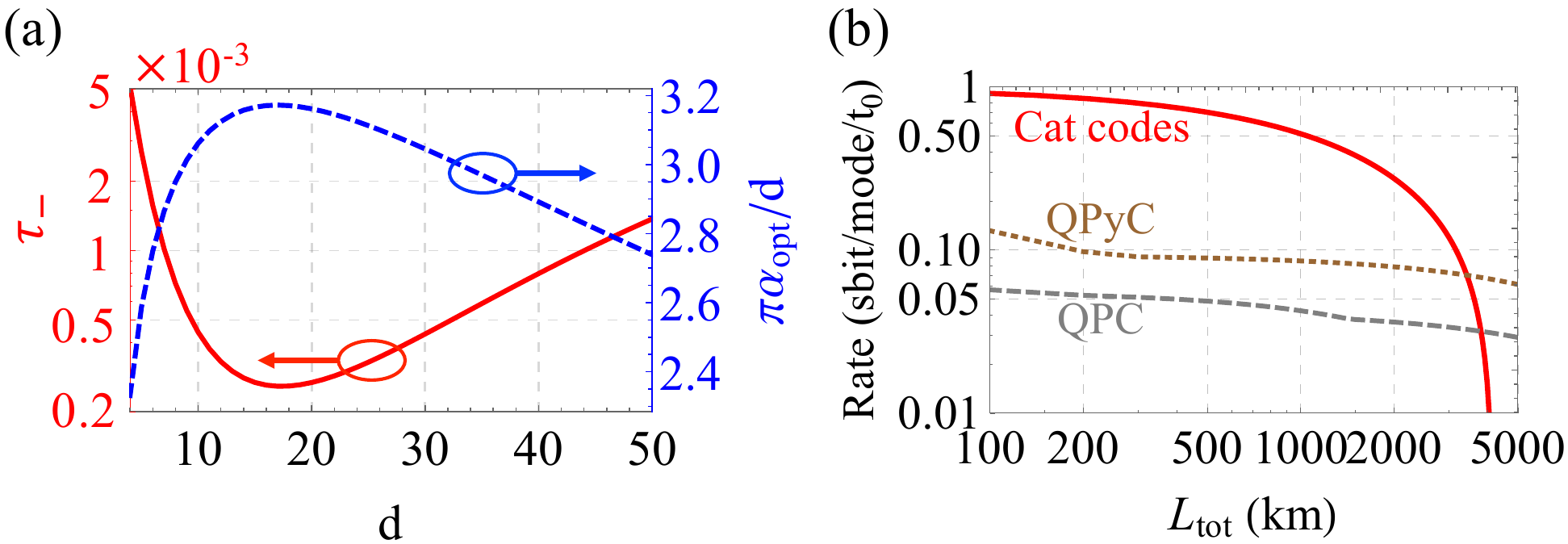}

\protect\caption{Optimized performance of cat codes for QRs with $\eta=99.5\%$ and
comparison with selected DV schemes. (a). Minimum effective error
rate $\tau_{-}$ (red) and associated optimum arc length $\frac{\pi\alpha_{\mathrm{opt}}}{d}$
(blue). (b). Optimized SKRPM over long distances for one-way QRs with
cat codes (red solid), quantum polynomial codes \cite{Muralidharan2015}
(brown dotted) and quantum parity code \cite{Muralidharan2014,Muralidharan2016}
(gray dashed). $t_{0}$ is the gate operation time taken as the same
for three schemes. \label{fig: optimization over long distance with QRs}}
\end{figure}

We introduce intermediate repeater stations with a small spacing $L_{0}$
($\ll L_{\mathrm{att}}$), so that the fiber attenuation induced loss
errors are correctable. Given near-unity coupling efficiency $\eta$
we have $\gamma\approx\tilde{L}_{0}+2\left(1-\eta\right)$, with $\tilde{L}_{0}=L_{0}/L_{\mathrm{att}}$
for the dimensionless repeater spacing. The goal is to minimize the
effective error rate
\begin{equation}
\tau_{-}\left(\alpha,\ \tilde{L}_{0},\ d\right)=\Gamma_{-}\left(\alpha,\ \gamma,\ d\right)/\tilde{L}_{0}.
\end{equation}
Fig.~\ref{fig: optimization over long distance with QRs}(a) shows
the minimized effective error rate as a function of $d$ for $\eta=99.5\%$
with the corresponding optimized arc length between neighboring coherent
states $\pi\alpha_{\mathrm{opt}}/d$. Note that the minimized error
rate is anti-correlated with the arc length $\pi\alpha_{\mathrm{opt}}/d$,
because increasing arc length suppresses the coherent component overlap
and consequently reduces the back-action induced dephasing. For small
$d$, the overall bit-flip error can be better suppressed by increasing
$d$ to correct more excitation loss errors; for large $d$, however,
the typical number of excitation losses is $\gamma\alpha^{2}\propto\gamma d^{2}$,
which will exceed the capability of QEC. Hence, there is an optimized
choice of $d$ that minimizes the overall error.

For one-way QRs with cat codes, the entire repeater chain can be characterized
by
\begin{equation}
\mathcal{E}^{N}=\left(\mathcal{R}\circ\mathcal{L}\right)^{N},
\end{equation}
with $N=L_{\mathrm{tot}}/\left(\tilde{L}_{0}^{\mathrm{opt}}L_{\mathrm{att}}\right)$
intermediate repeater stations. More specifically, we consider a four-state
quantum key distribution protocol. With $\mathcal{E}^{N}$ written
as a $4\times4$ matrix with entries $\mathcal{E}_{ij}^{N}$, quantum
bit error rates in the Z- and X-basis can be derived as $Q_{Z}=\left(\mathcal{E}_{14}^{N}+\mathcal{E}_{41}^{N}\right)/2$
and $Q_{X}=\left[\left(\mathcal{E}_{11}^{N}+\mathcal{E}_{14}^{N}+\mathcal{E}_{41}^{N}+\mathcal{E}_{44}^{N}\right)-\left(\mathcal{E}_{22}^{N}+\mathcal{E}_{23}^{N}+\mathcal{E}_{32}^{N}+\mathcal{E}_{33}^{N}\right)\right]/4$
\footnote{In the numerical calculations, we use these expressions that are valid
for a general qubit channel. It can be shown that, with approximations
in ref.~\cite{SupplementalMaterial}, for $N=1$, $Q_{Z}\approx\epsilon_{f}$
and $Q_{X}\approx\epsilon_{d}$.%
}, respectively.  Since using multiple modes may carry a large resource
overhead, here we use the  secure key rate per mode (SKRPM) to evaluate
the performance of one-way QRs \cite{Namiki2016}. For single-mode
encoding schemes, the SKRPM is $\mathrm{Rate}=1-H\left(Q_{Z}\right)-H\left(Q_{X}\right)$,
where $H\left(p\right)=-p\log_{2}p-\left(1-p\right)\log_{2}\left(1-p\right)$
is the binary entropy function \cite{Wang2005}. Fig.~\ref{fig: optimization over long distance with QRs}(b)
shows the optimized SKRPM for one-way QRs with cat codes with $\eta=99.5\%$
for distributing quantum keys over long distances and, in comparison,
the optimized SKRPM for multi-mode DV encoding quantum parity code
(QPC) \cite{Muralidharan2014,Muralidharan2016} and quantum polynomial
code (QPyC) \cite{Muralidharan2015}. With high coupling efficiency,
as a single-mode encoding, cat codes outperform conventional DV quantum
codes due to efficient use of the bosonic mode.

\paragraph*{Conclusion and outlook.}

We have investigated cat codes for protecting quantum states against
bosonic excitation loss. At the encoded level, there are two major
types of uncorrectable errors, logical bit-flip error, due to excessive
excitation loss and logical dephasing error, induced by back-action.
We have demonstrated that non-trivial combinations of coherent amplitude
and logical subspace can efficiently suppress logical dephasing error,
and lead to significantly improved QEC performance. We expect this
feature of suppressed back-action from the environment to be observed
for other approximate CV quantum codes as $\left\langle 0_{L}\right|a^{\dagger}a\left|0_{L}\right\rangle =\left\langle 1_{L}\right|a^{\dagger}a\left|1_{L}\right\rangle $
is satisfied and the balance between the back-action and excessive
excitation loss could be useful for the optimization of their QEC
capabilities. Comparison between cat codes and other known single-mode
schemes, such as GKP codes \cite{Gottesman2001,Terhal2016,Terhal2016a}
and binomial codes \cite{Michael2016}, over LBC could shed further
light on the optimal construction of single-mode CV encodings. We
notice that cat codes become less favorable, compared with conventional
multi-mode schemes, in case of long communication distance (Fig.~\ref{fig: optimization over long distance with QRs}(b))
or high coupling loss \cite{SupplementalMaterial}, as a result of
high occupation of a single bosonic mode. This can motivate us to
explore unconventional multi-mode CV encodings with multiple excitations
per mode \cite{Harrington2001} that may asymptotically achieve the
channel capacity of LBC. 

As an application, we have explored one-way quantum communication
over long distances with cat codes and found that given high-fidelity
coupling into and out of the repeaters this single-mode continuous
variable scheme can outperform conventional schemes with single excitation
occupying multiple modes in terms of secure key rate per mode. With
recent developments of efficient coupling between fiber and optical
waveguide \cite{Tiecke2015}, and high-fidelity frequency conversion
between optical and microwave modes \cite{Andrews2014,OBrien2014,Zou2016},
we may envision realistic quantum repeaters consisting of superconducting
circuits for error correction and optical-microwave quantum transducers
to protect transmitted quantum information against optical loss in
fiber channels. 

We thank Kasper Duivenvoorden, Jungsang Kim, Norbert Lütkenhaus, Marios
H. Michael, Ananda Roy, Chao Shen, Barbara Terhal, Hong Tang for stimulating
discussions. We acknowledge support from the ARL-CDQI, ARO (W911NF-14-1-0011,
W911NF-14-1-0563), ARO MURI (W911NF-16-1-0349), NSF (DMR-1609326,
DGE-1122492), AFOSR MURI (FA9550-14-1- 0052, FA9550-14-1-0015), Alfred
P. Sloan Foundation (BR2013-049), and Packard Foundation (2013-39273).

\emph{Note added}: During the preparation of the manuscript, the authors
became aware of a related work on cat codes \cite{Bergmann2016}.
Different from that work, here we have proposed a deterministic amplitude
restoration for QEC recovery and investigated combined optimization
of amplitude and logical subspace.

\bibliographystyle{apsrev4-1}
\bibliography{cat_ref}

\clearpage{}\newpage{}

\newpage{}

\onecolumngrid
\renewcommand{\thefigure}{S\arabic{figure}}
\setcounter{figure}{0} 
\renewcommand{\thepage}{S\arabic{page}}
\setcounter{page}{0} 
\renewcommand{\theequation}{S.\arabic{equation}}
\setcounter{equation}{0} 
\setcounter{section}{0}

\begin{center}
\textbf{\textsc{\LARGE{}Supplementary Material}}
\par\end{center}{\LARGE \par}

\section{Analysis of quantum channel $\mathcal{E}$ and diamond norm $\left\Vert \mathcal{E}-\mathcal{I}\right\Vert _{\diamond}$
\label{sec: Appednix.-A-Eigenvalue}}

In the following, we analytically show that $\mathcal{E}$ can be
approximated as a Pauli channel and calculate the diamond norm $\left\Vert \mathcal{E}-\mathcal{I}\right\Vert _{\diamond}$.
We begin by specifying two assumptions that are used throughout the
analysis
\begin{enumerate}
\item \label{enu:Approx-1}$\epsilon_{f}=\stackrel[k=0]{d-1}{\sum}T_{k+d}=\mathcal{O}\left(\frac{e^{-\Delta}\Delta^{d}}{d!}\right)\ll1$,
which physically implies that bit-flip error is considerably small.
\item \label{enu:Approx-2}Defining $\delta=\left|\Braket{\alpha|\alpha e^{i\frac{\pi}{d}}}\right|=\left|\exp\left(\alpha^{2}e^{i\frac{\pi}{d}}-\alpha^{2}\right)\right|$,
then $\delta\approx e^{-\frac{1}{2}\left(\frac{\pi\alpha}{d}\right)^{2}}\ll1$,
which physically implies that the overlap between neighboring coherent
states is considerably small.
\end{enumerate}
Then, the normalization factor
\begin{eqnarray}
\mathcal{N}_{n}\left(\alpha\right) & = & \stackrel[k=0]{2d-1}{\sum}\omega^{-kn}\exp\left[\left(\omega^{k}-1\right)\alpha^{2}\right]\nonumber \\
 & = & \stackrel[j=0]{2d-1}{\sum}e^{-2\alpha^{2}\sin^{2}\frac{j\pi}{2d}}\cos\left(\frac{jn\pi}{d}-\alpha^{2}\sin\frac{j\pi}{d}\right)\nonumber \\
 & = & 1+\zeta_{n}\left(\alpha\right)+\mathcal{O}\left(\delta^{4}\right),\label{eq: normaliztion factor}
\end{eqnarray}
where $\zeta_{n}\left(\alpha\right)=2e^{-2\alpha^{2}\sin^{2}\frac{\pi}{2d}}\cos\left(\frac{n\pi}{d}-\alpha^{2}\sin\frac{\pi}{d}\right)=\mathcal{O}\left(\delta\right)$.
With Approx. \ref{enu:Approx-1}-\ref{enu:Approx-2} and Eq.~(\ref{eq: normaliztion factor}),
we note 
\begin{eqnarray*}
\stackrel[k=0]{d-1}{\sum}T_{k+d}C_{sk}^{2} & = & \stackrel[k=0]{d-1}{\sum}T_{k+d}\frac{\mathcal{N}_{s-k}\left(\alpha'\right)}{\mathcal{N}_{d+s}\left(\alpha\right)}\\
 & = & \frac{\stackrel[k=0]{d-1}{\sum}T_{k+d}+\stackrel[k=0]{d-1}{\sum}T_{k+d}\left[\mathcal{N}_{s-k}\left(\alpha'\right)-1\right]}{1+\left[\mathcal{N}_{d+s}\left(\alpha\right)-1\right]}\\
 & = & \frac{\stackrel[k=0]{d-1}{\sum}T_{k+d}+\stackrel[k=0]{d-1}{\sum}T_{k+d}\mathcal{O}\left(\delta\right)}{1+\mathcal{O}\left(\delta\right)}\\
 & = & \epsilon_{f}+\mathcal{O}\left(\epsilon_{f}\delta\right).
\end{eqnarray*}
Similarly, $\stackrel[k=0]{d-1}{\sum}T_{k+d}B_{sk}C_{sk}=\epsilon_{f}+\mathcal{O}\left(\epsilon_{f}\delta\right)$
and $\stackrel[k=0]{d-1}{\sum}T_{k+d}B_{sk}^{2}=\epsilon_{f}+\mathcal{O}\left(\epsilon_{f}\delta\right)$.
Therefore, we may simplify $\mathcal{E}$ as
\begin{eqnarray}
\mathcal{E} & = & \left(\begin{array}{cccc}
\lambda_{1} & 0 & 0 & \epsilon_{f}\\
0 & \lambda_{2} & \epsilon_{f} & 0\\
0 & \epsilon_{f} & \lambda_{2} & 0\\
\epsilon_{f} & 0 & 0 & \lambda_{3}
\end{array}\right)+\mathcal{O}\left(\epsilon_{f}\delta\right),\label{eq: approximated tranmission matrix}
\end{eqnarray}
where $\lambda_{1}=\stackrel[k=0]{d-1}{\sum}T_{k}A_{sk}^{2}$, $\lambda_{2}=\stackrel[k=0]{d-1}{\sum}T_{k}A_{sk}D_{sk}$
and $\lambda_{3}=\stackrel[k=0]{d-1}{\sum}T_{k}D_{sk}^{2}$, and the
back-action coefficients are $A_{sk}=\sqrt{\mathcal{N}_{s-k}\left(\alpha'\right)/\mathcal{N}_{s}\left(\alpha\right)}$,
$B_{sk}=\sqrt{\mathcal{N}_{d+s-k}\left(\alpha'\right)/\mathcal{N}_{s}\left(\alpha\right)}$,
$C_{sk}=\sqrt{\mathcal{N}_{s-k}\left(\alpha'\right)/\mathcal{N}_{d+s}\left(\alpha\right)}$
and $D_{sk}=\sqrt{\mathcal{N}_{d+s-k}\left(\alpha'\right)/\mathcal{N}_{d+s}\left(\alpha\right)}$.

\subsection{$\lambda_{1}$ and $\lambda_{3}$ }

First of all, we rewrite the expression of $\lambda_{1}$ as
\begin{eqnarray}
1-\lambda_{1} & = & 1-\frac{\stackrel[k=0]{d-1}{\sum}T_{k}\mathcal{N}_{s-k}\left(\alpha'\right)}{\mathcal{N}_{s}\left(\alpha\right)}\nonumber \\
 & = & 1-\frac{\stackrel[k=0]{d-1}{\sum}T_{k}+\stackrel[k=0]{d-1}{\sum}T_{k}\left[\mathcal{N}_{s-k}\left(\alpha'\right)-1\right]}{1+\left[\mathcal{N}_{s}\left(\alpha\right)-1\right]}.
\end{eqnarray}

Since $T_{k}=\stackrel[m=0]{\infty}{\sum}\frac{e^{-\Delta}\Delta^{2md+k}}{\left(2md+k\right)!}$
and $\epsilon_{f}=1-\stackrel[k=0]{d-1}{\sum}T_{k}$, we have 
\begin{eqnarray}
\stackrel[k=0]{d-1}{\sum}T_{k}\left[\mathcal{N}_{s-k}\left(\alpha'\right)-1\right] & = & \stackrel[k=0]{d-1}{\sum}\stackrel[m=0]{\infty}{\sum}\frac{e^{-\Delta}\Delta^{2md+k}}{\left(2md+k\right)!}\left[\mathcal{N}_{s-k}\left(\alpha'\right)-1\right]\nonumber \\
 & = & \stackrel[k=0]{\infty}{\sum}\frac{e^{-\Delta}\Delta^{k}}{k!}\stackrel[j=1]{2d-1}{\sum}e^{-2\alpha'^{2}\sin^{2}\frac{j\pi}{2d}}\cos\left[\frac{j\left(s-k\right)\pi}{d}-\alpha'^{2}\sin\frac{j\pi}{d}\right]+\mathcal{O}\left(\epsilon_{f}\delta\right)\nonumber \\
 & = & \stackrel[j=1]{2d-1}{\sum}e^{\alpha'^{2}\cos\frac{j\pi}{d}-\alpha^{2}}\stackrel[k=0]{\infty}{\sum}\frac{\Delta^{k}}{k!}\cos\left[\frac{j\left(s-k\right)\pi}{d}-\alpha'^{2}\sin\frac{j\pi}{d}\right]+\mathcal{O}\left(\epsilon_{f}\delta\right)\nonumber \\
 & = & \stackrel[j=1]{2d-1}{\sum}e^{\alpha'^{2}\cos\frac{j\pi}{d}-\alpha^{2}}e^{\Delta\cos\frac{j\pi}{d}}\cos\left(\frac{js\pi}{d}-\alpha^{2}\sin\frac{j\pi}{d}\right)+\mathcal{O}\left(\epsilon_{f}\delta\right)\nonumber \\
 & = & \mathcal{N}_{s}\left(\alpha\right)-1+\mathcal{O}\left(\epsilon_{f}\delta\right),\label{eq:sumapprox}
\end{eqnarray}
where we use the equality 
\begin{eqnarray*}
\stackrel[k=0]{\infty}{\sum}\frac{\Delta^{k}}{k!}\cos\left[\frac{j\left(s-k\right)\pi}{d}-\alpha'^{2}\sin\frac{j\pi}{d}\right] & = & e^{\Delta\cos\frac{j\pi}{d}}\cos\left(\frac{js\pi}{d}-\alpha^{2}\sin\frac{j\pi}{d}\right).
\end{eqnarray*}

Therefore, we arrive at 
\begin{eqnarray}
1-\lambda_{1} & = & \frac{1+\left[\mathcal{N}_{s}\left(\alpha\right)-1\right]-\stackrel[k=0]{d-1}{\sum}T_{k}-\stackrel[k=0]{d-1}{\sum}T_{k}\left[\mathcal{N}_{s-k}\left(\alpha'\right)-1\right]}{1+\left[\mathcal{N}_{s}\left(\alpha\right)-1\right]}\nonumber \\
 & = & \frac{\epsilon_{f}+\mathcal{O}\left(\epsilon_{f}\delta\right)}{1+\mathcal{O}\left(\delta\right)}\nonumber \\
 & = & \epsilon_{f}+\mathcal{O}\left(\epsilon_{f}\delta\right)
\end{eqnarray}

Similarly we can also obtain 
\begin{equation}
1-\lambda_{3}=\epsilon_{f}+\mathcal{O}\left(\epsilon_{f}\delta\right).
\end{equation}

\subsection{$\lambda_{2}$ }

The expression of $\lambda_{2}$ is 
\begin{eqnarray}
1-\lambda_{2} & = & 1-\frac{\stackrel[k=0]{d-1}{\sum}T_{k}\sqrt{\mathcal{N}_{s-k}\left(\alpha'\right)\mathcal{N}_{d+s-k}\left(\alpha'\right)}}{\sqrt{\mathcal{N}_{s}\left(\alpha\right)\mathcal{N}_{d+s}\left(\alpha\right)}}.
\end{eqnarray}

From Eq.~(\ref{eq: normaliztion factor}) we know that 
\begin{eqnarray*}
\mathcal{N}_{s}\left(\alpha\right) & = & 1+\zeta_{s}\left(\alpha\right)+\mathcal{O}\left(\delta^{4}\right),\\
\mathcal{N}_{d+s}\left(\alpha\right) & = & 1-\zeta_{s}\left(\alpha\right)+\mathcal{O}\left(\delta^{4}\right),
\end{eqnarray*}
and
\begin{eqnarray}
\sqrt{\mathcal{N}_{s}\left(\alpha\right)\mathcal{N}_{d+s}\left(\alpha\right)} & = & 1-\frac{\zeta_{s}\left(\alpha\right)^{2}}{2}+\mathcal{O}\left(\delta^{4}\right)
\end{eqnarray}
we have
\begin{eqnarray}
1-\lambda_{2} & = & 1-\frac{\stackrel[k=0]{d-1}{\sum}T_{k}\left[1-\frac{\zeta_{s-k}\left(\alpha'\right)^{2}}{2}+\mathcal{O}\left(\delta^{4}\right)\right]}{1-\frac{\zeta_{s}\left(\alpha\right)^{2}}{2}+\mathcal{O}\left(\delta^{4}\right)}\nonumber \\
 & = & \frac{\epsilon_{f}-\frac{\zeta_{s}\left(\alpha\right)^{2}}{2}+\mathcal{O}\left(\delta^{4}\right)+\stackrel[k=0]{d-1}{\sum}T_{k}\left[\frac{\zeta_{s-k}\left(\alpha'\right)^{2}}{2}+\mathcal{O}\left(\delta^{4}\right)\right]}{1-\frac{\zeta_{s}\left(\alpha\right)^{2}}{2}+\mathcal{O}\left(\delta^{4}\right)}.\label{eq:lambda2_0}
\end{eqnarray}
Similar to the approximation in Eq.~(\ref{eq:sumapprox}), we have
\begin{eqnarray}
\stackrel[k=0]{d-1}{\sum}T_{k}\frac{\zeta_{s-k}\left(\alpha'\right)^{2}}{2} & = & 2e^{-4\alpha'^{2}\sin^{2}\frac{\pi}{2d}}\stackrel[k=0]{\infty}{\sum}\frac{e^{-\Delta}\Delta^{k}}{k!}\cos^{2}\left[\frac{\left(s-k\right)\pi}{d}-\alpha'^{2}\sin\frac{\pi}{d}\right]+\mathcal{O}\left(\epsilon_{f}\delta^{2}\right)\nonumber \\
 & = & e^{-4\alpha'^{2}\sin^{2}\frac{\pi}{2d}}\left\{ 1+e^{-2\Delta\sin^{2}\frac{\pi}{d}}\cos\left[\frac{2s\pi}{d}-2\alpha'^{2}\sin\frac{\pi}{d}-\Delta\sin\frac{2\pi}{d}\right]\right\} +\mathcal{O}\left(\epsilon_{f}\delta^{2}\right),
\end{eqnarray}
and hence 
\begin{eqnarray*}
-\frac{\zeta_{s}\left(\alpha\right)^{2}}{2}+\stackrel[k=0]{d-1}{\sum}T_{k}\frac{\zeta_{s+k}\left(\alpha'\right)^{2}}{2} & = & e^{-4\alpha^{2}\sin^{2}\frac{\pi}{2d}}\left\{ e^{4\Delta\sin^{2}\frac{\pi}{2d}}-1-\left[\cos\Phi-e^{\mu}\cos\left(\Phi+\psi\right)\right]\right\} +\mathcal{O}\left(\epsilon_{f}\delta^{2}\right)\\
 & = & e^{-4\alpha^{2}\sin^{2}\frac{\pi}{2d}}\left[e^{4\Delta\sin^{2}\frac{\pi}{2d}}-1-\left(1-2e^{\mu}\cos\psi+e^{2\mu}\right)^{\frac{1}{2}}\cos\left(\Phi+\varphi\right)\right]+\mathcal{O}\left(\epsilon_{f}\delta^{2}\right)
\end{eqnarray*}
where we denote $\mu=2\Delta\left(2\sin^{2}\frac{\pi}{2d}-\sin^{2}\frac{\pi}{d}\right)$,
$\psi=\Delta(2\sin\frac{\pi}{d}-\sin\frac{2\pi}{d})$, $\Phi=\frac{2s\pi}{d}-2\alpha^{2}\sin\frac{\pi}{d}$
and $\varphi=\arctan\frac{e^{\mu}\sin\psi}{1-e^{\mu}\cos\psi}$. Therefore
\begin{eqnarray}
1-\lambda_{2} & = & \epsilon_{f}+e^{-4\alpha^{2}\sin^{2}\frac{\pi}{2d}}\left[e^{4\Delta\sin^{2}\frac{\pi}{2d}}-1-\left(1-2e^{\mu}\cos\psi+e^{2\mu}\right)^{\frac{1}{2}}\cos\left(\Phi+\varphi\right)\right]+\mathcal{O}\left(\epsilon_{f}\delta^{2}\right).\label{eq:final approximation of lambda2}\\
 & \equiv & \epsilon_{f}+2\epsilon_{d}+\mathcal{O}\left(\epsilon_{f}\delta^{2}\right).\nonumber 
\end{eqnarray}
Plugging the analytical expressions of $\lambda_{n}\ (n=1,2,3)$ into
Eq.~(\ref{eq: approximated tranmission matrix}), we can approximate
$\mathcal{E}$ as a qubit Pauli channel as in Eq.~(\ref{eq: quantum channel}).

\section{Quantification of decoherence suppression}

We quantify the improvement in suppressing decoherence using our approach
by further simplifying $\Gamma_{-}=\epsilon_{f}+\frac{1}{2}e^{-4\alpha^{2}\sin^{2}\frac{\pi}{2d}}\left[e^{4\Delta\sin^{2}\frac{\pi}{2d}}-1-\left(1-2e^{\mu}\cos\psi+e^{2\mu}\right)^{\frac{1}{2}}\right]$.
Considering $\frac{\pi}{d}\ll1$, we shall approximate
\begin{eqnarray*}
e^{4\Delta\sin^{2}\frac{\pi}{2d}}-1 & = & \Delta\left(\frac{\pi}{d}\right)^{2}\left[1+\frac{1}{2}\left(\Delta\left(\frac{\pi}{d}\right)^{2}-\frac{1}{6}\right)\left(\frac{\pi}{d}\right)^{2}\right]+\mathcal{O}\left(\frac{\pi}{d}\right)^{6}\\
\left(1-2e^{\mu}\cos\psi+e^{2\mu}\right)^{\frac{1}{2}} & = & \Delta\left(\frac{\pi}{d}\right)^{2}\left[1-\frac{1}{2}\left(\Delta\left(\frac{\pi}{d}\right)^{2}+\frac{1}{6}\right)\left(\frac{\pi}{d}\right)^{2}\right]+\mathcal{O}\left(\frac{\pi}{d}\right)^{6},
\end{eqnarray*}
 and hence 
\begin{equation}
\Gamma_{-}=\epsilon_{f}+\frac{1}{2}e^{-4\alpha^{2}\sin^{2}\frac{\pi}{2d}}\left[\Delta^{2}\left(\frac{\pi}{d}\right)^{4}+\mathcal{O}\left(\frac{\pi}{d}\right)^{6}\right].\label{eq: approximated Tau_}
\end{equation}
On the other hand, 
\begin{eqnarray}
\bar{\Gamma} & = & \epsilon_{f}+\frac{1}{2}e^{-4\alpha^{2}\sin^{2}\frac{\pi}{2d}}\left(e^{4\Delta\sin^{2}\frac{\pi}{2d}}-1\right)\nonumber \\
 & = & \epsilon_{f}+\frac{1}{2}e^{-4\alpha^{2}\sin^{2}\frac{\pi}{2d}}\left[\Delta\left(\frac{\pi}{d}\right)^{2}+\mathcal{O}\left(\frac{\pi}{d}\right)^{4}\right].\label{eq: approximated Taubar}
\end{eqnarray}
At the regime where back-action induced dephasing dominates, we can
see from Eq.~(\ref{eq: approximated Tau_}-\ref{eq: approximated Taubar})
that the decoherence is reduced from $\mathcal{O}\left(\Delta\pi^{2}/d^{2}\right)$
to $\mathcal{O}\left[\left(\Delta\pi^{2}/d^{2}\right)^{2}\right]$.
\begin{figure}
\includegraphics[scale=0.7]{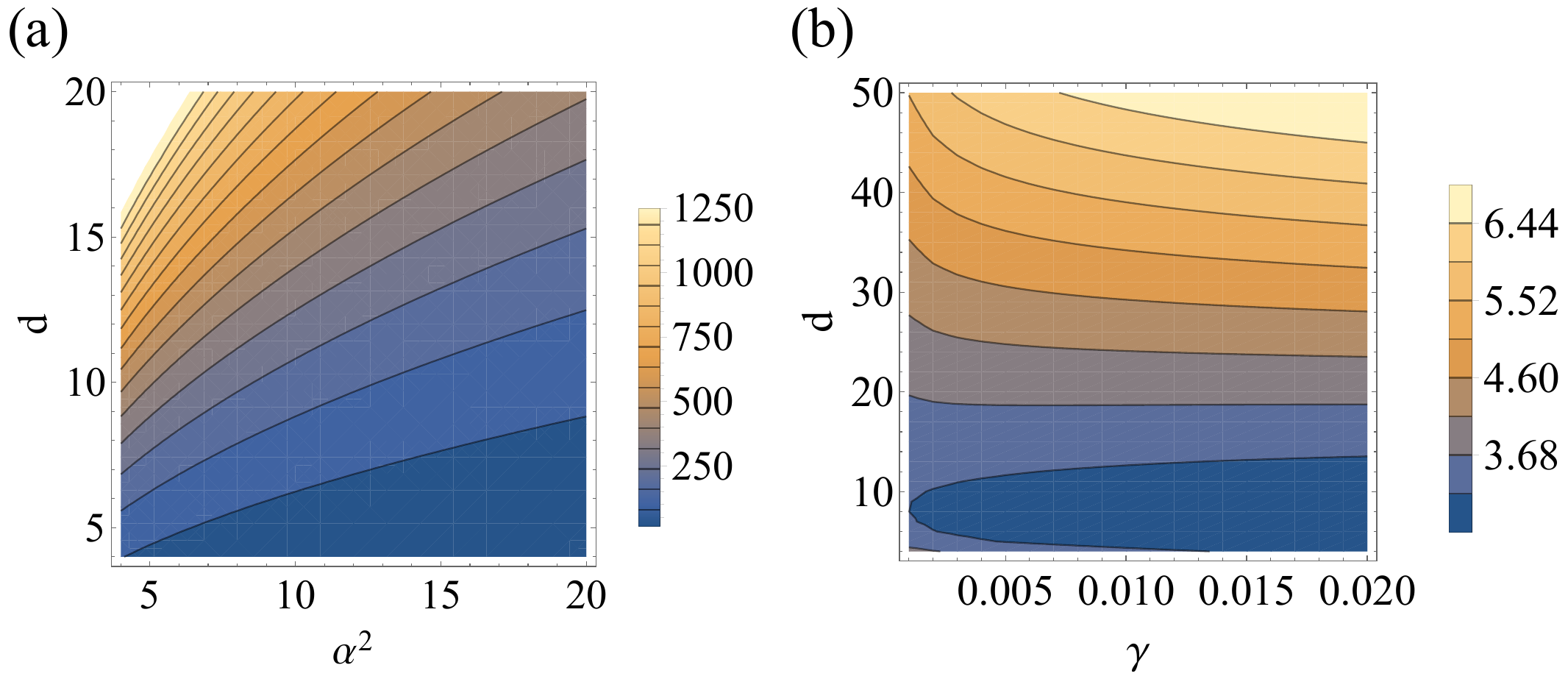}

\protect\caption{Improvement in suppressing decoherence using cat codes with the proposed
recovery. (a). $\Delta\pi^{2}/d^{2}$ with $\gamma=0.005$. (b). $\bar{\Gamma}\left(\alpha_{subo},\gamma,d\right)/\Gamma_{-}\left(\alpha_{o},\gamma,d\right)$\label{fig:Improvement in suppressing decoherence}}
\end{figure}
In Fig.~\ref{fig:Improvement in suppressing decoherence}(a) we show
$\frac{1}{\Delta\pi^{2}/d^{2}}$ with fixed $\gamma=0.005$, which
demonstrates that, by incorporating our recovery, in the small $\alpha$
regime that is mostly relevant, the coherence duration of encoded
states can be improved by orders of magnitude leading to substantial
extension of the lifetime of cavity-based quantum memories and secure
communication rate for quantum communication. 

In the following, we look at the overall picture and investigate how
much our approach at $\alpha_{o}$ outperforms the otherwise best
strategy, which is to pick the $\alpha$ (denoted as $\alpha_{subo}$)
corresponding to the crossing between $\frac{1}{2}e^{-4\alpha^{2}\sin^{2}\frac{\pi}{2d}}\left(e^{4\Delta\sin^{2}\frac{\pi}{2d}}-1\right)$
and $\epsilon_{f}$. We evaluate the ratio of $\bar{\Gamma}\left(\alpha_{subo},\gamma,d\right)$
and $\Gamma_{-}\left(\alpha_{o},\gamma,d\right)$ in Fig.~\ref{fig:Improvement in suppressing decoherence}(b)
and observe a $3\sim6$ times improvement in suppressing decoherence,
depending on $d$ and $\gamma$. Noting that $\alpha_{subo}$ is always
larger than $\alpha_{o}$, our approach works better in terms of both
performance and feasibility.

\section{Repetitive correction with cat codes}

We consider repetitive QEC with cat codes, the goal of which is to
further extend the coherence duration of encoded quantum states to
a total distance $L_{tot}$ for quantum communication or a total time
of $T_{tot}$ for localized quantum memories. In this case the waiting
period before each recovery, or equivalently the total number of recoveries,
is also upon optimization. we consider quantum communication with
QRs to showcase and optimize the effective error rate $\tau_{-}\left(\alpha,\tilde{L}_{0},d\right)=\Gamma_{-}\left(\alpha,\gamma,d\right)/\tilde{L}_{0}$
where $\tilde{L}_{0}$ is the dimensionless repeater spacing.

In Fig.~\ref{fig: L0, epsilon_f and different couplings}(a), we
show the optimized spacing $\tilde{L}_{0}^{opt}$ and associated bit-flip
error rate $\epsilon_{f}$. We can see that $\tilde{L}_{0}^{opt}$
accounting for the loss induced by transmission is always smaller
than the coupling loss $2\left(1-\eta\right)$ which is $0.01$ in
this case. Therefore, we may approximately neglect how changing $\tilde{L}_{0}^{opt}$
affects $\gamma$ and only consider its effect on the total number
of stations $N=\frac{L_{tot}}{\tilde{L}_{0}^{opt}L_{att}}$. In Fig.~\ref{fig: L0, epsilon_f and different couplings}(b),
the optimized secure key rates (per mode) over long distances for
$\eta=99.4\%-99.6\%$ are shown. We can see that the performance of
cat codes is sensitive to coupling efficiency due to the fact that
the scheme uses only one mode. Nonetheless, moderate coupling efficiency,
such as $\eta=99.5\%$, is already good for communication over $L_{tot}\approx10^{3}\mathrm{km}$
and small improvement in $\eta$ can considerably increase the rate.
\begin{figure}[h]
\includegraphics[scale=0.7]{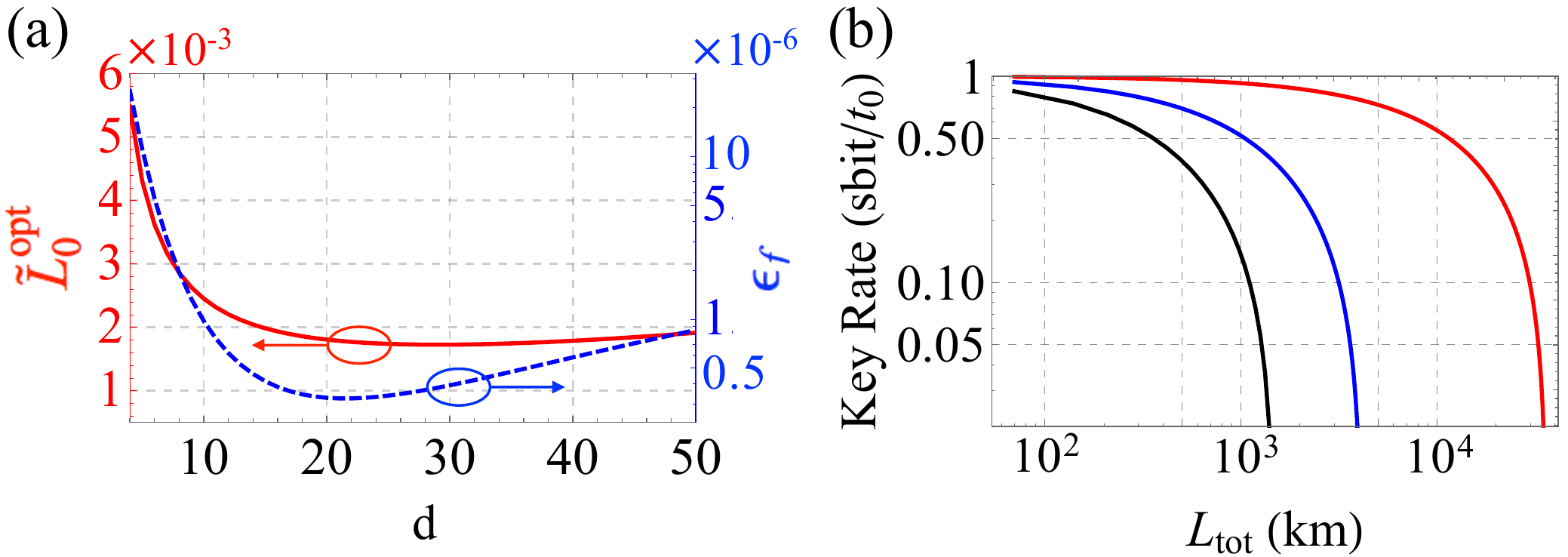}

\protect\caption{(a). Optimized spacing $\tilde{L}_{0,opt}$ (red) and associated bit-flip
error rate per transmission $\epsilon_{f}$ (blue) with $\gamma=0.005$.
(b). Optimized secure key rate with $\eta=99.4\%$ (black), $99.5\%$
(blue) and $99.6\%$ (red). \label{fig: L0, epsilon_f and different couplings}}
\end{figure}

\end{document}